\documentstyle[11pt]{article}
\newcommand{\be}{\begin{equation}}
\newcommand{\ee}{\end{equation}}

\newcommand{\ts}{\hspace{3pt}}
\newcommand{\bea}{\begin{eqnarray}}
\newcommand{\eea}{\end{eqnarray}}

\begin{document} \noindent
Reproduced from\\ FOUNDATIONS OF PHYSICS
Vol.\ts 3, No.1, March 1973, p.\ts 109--116\\
(with several corrections and reformulations)

\vspace*{2.0cm}

\noindent
{\bf  TOWARD A QUANTUM THEORY OF OBSERVATION	}

\vskip 1.cm
\begin{quote}
\noindent
{\bf H. D. Zeh}
\vskip 0.2cm
\noindent
Institut f\"ur Theoretische Physik\\ Universit\"at Heidelberg\\ www.zeh-hd.de
\\

\end{quote}
\vskip .5cm

\noindent
{\bf Abstract:} The program of a physical concept of  information is 
outlined in the
framework of  quantum theory. A proposal is made for how to avoid the
introduction of axiomatic observables. The conventional (collapse) 
and the Everett
interpretations of quantum theory may in principle lead to different dynamical
consequences. Finally, a formal ensemble description not based on a concept of
lacking information is discussed.

\vskip2.0cm

\noindent
{\bf 1. INTRODUCTION}  \vskip.5cm

\noindent
In the
conventional interpretation of quantum theory, the wave function or 
state vector
describing a physical system evolves in time according to two 
different laws: (a) the
Schr\"odinger equation, being deterministic in the sense that the 
wave function is
determined completely if given at a certain time, and (b) the ``reduction" or
``collapse", which describes the transition of the wave function into 
one member of a
set of mutually orthogonal states, this set (defining an ``observable" up to
scale transformations) being phenomenologically associated with a 
certain type of
measurement.

The second kind of time evolution has to be taken into account
regardless of whether it is called ``real" or ``due to the increase 
of information",
in particular since communication is to be considered a physical process.

All attempts to derive the reduction from a Schr\"odinger equation 
for  the total
system including the measurement apparatus have been shown to be doomed to
failure [1,2]. Derivations from a master equation seem to be more 
appropriate but lead
to (apparently related) difficulties in deriving the master equation 
itself [3]. In
both cases similar pseudo-arguments are often introduced by using an 
abstract {\it a
priori} concept of information instead of a physical one.

Everett proposed an
interpretation of quantum theory that avoids the reduction of the 
wave function, and
instead postulates the universal validity of the Schr\"odinger 
equation [4]. This
means that the ``other" components (which would disappear according 
to the reduction
postulate) still ``exist" after a measurement. However, they are 
correlated with
different states of the apparatus and subsequently with different 
states of human
observers. Therefore Everett has to assume that ``we are aware" of 
only one of these
world components, and that the world apparently ``splits" into many components
whenever a measurement-like process occurs.

Is it not entirely meaningless to postulate the
existence of world components that are not observed? Even if the 
latter were true,
it would at least be conventional to regard those things as 
``existing in reality"
which are extrapolated from the observed by means of established 
laws. It will be
demonstrated in Section 3, however, that the different world 
components {\it may}
interact if the total wave function obeys a Schr\"odinger equation -- 
as assumed by
Everett. Consequently, in spite of the linearity of the equation of 
motion, the time
evolution of the ``observed component" might show a dependence on the other
components.

Proposals to avoid the reduction by
considering several successive measurements as {\it one} correlation 
experiment [5]
either leave the process of ``preparation" of the initial state conceptually
unexplained or have to refer to a universal wave function again. In 
the latter case,
all measurements (in the widest sense) ever performed must be 
considered as one great
correlation experiment (that is still  going on). None of these interpretations
describes the complete process of observation (including the increase 
of information)
{\it by means of a Schr\"odinger equation}.
\eject

\vskip.5cm \noindent {\bf 2. PHYSICAL DETERMINATION OF OBSERVABLES} \vskip.5cm

\noindent The Everett
interpretation in itself is not complete.  There are many possibilities of
decomposing a wave function into components
\be
\psi = \sum_i \phi_i \Phi^{(i)}
\ee
   where
$\phi_i$ are orthogonal states of a considered system, while the 
$\Phi^{(i)}$ are the
corresponding  ``relative states" of the ``remainder of the universe". Everett
suggested a decomposition with respect to different ``memory states" without
specifying them physically. In analogy to superselection rules, this 
proposal thus
excludes a splitting into {\it superpositions} of different memory states. The
phenomenological selection of memory states corresponds to the 
axiomatic introduction
of observables in the conventional description of measurements.

An essential property of memory
states is their stability with respect to external perturbations, 
even though they
may in turn easily affect their environment [6]. (Consider pointer 
positions, books,
or different DNS chains.) This property will cause a strong correlation of
memory properties with the remainder of the world. Dynamical 
stability (``robustness")
is therefore suggestive as a criterion for states with respect to which the
universal wave function splits (or, alternatively, collapses).

Another
possibility of specifying the splitting is based on the fundamental 
subject-object
relation of observation. Assume the total system to be conceptually 
cut into two
subsystems, one containing the object, the other one the observer. 
Then there is an
essentially unique decomposition of the total wave function into a single
sum of orthonormal states for both subsystems (the ``Schmidt 
canonical form") [7],
\be
\psi(t) = \sum_i \sqrt{p_i(t)}\phi_i(t)\Phi_i(t) \quad .
\ee
For dynamical and statistical reasons, this decomposition tends to keep
different memory states  in different terms of the series, irrespective
of the cut's precise position. (In contrast to Bohr's epistemology, 
the cut does here
not have to be interpreted as one between quantum and classical concepts.)

An Everett branching in terms of
dynamically stable states has the disadvantage of being conceptually 
approximate,
while that based on the Schmidt representation depends on the position of the
cut. Nonetheless, the second version will be preferred in the 
following. (One knows
from the pragmatic rules of quantum theory that the cut has to be put 
``far enough"
away from the object, that is, close enough to the observer [8]. 
Therefore it may be
chosen to have on the observer's side only the physical carrier of 
his consciousness
-- perhaps the cerebral cortex or parts thereof -- thus keeping even 
his personal
memory in the ``outside world". If the cut were placed somewhere in 
between object
and observer, the Schmidt decomposition would be too fine-grained. A 
reduction to one
of the components of (2) would then describe a much greater increase of
information than that according to the axiomatic observables.)

\vskip.5cm \noindent
{\bf 3. CAN THE EVERETT INTERPRETATION BE VERIFIED?} \vskip.5cm

\noindent
Provided the wave function $\psi(t)$ of some total system  exists and obeys a
Schr\"odinger equation, the coefficients and components of expansion 
(2) vary in
time according to the nonlinear set of equations [9]
\bea
&&  \hbar {d\sqrt{p_i} \over dt} = {\rm Im} \langle \phi_i \Phi_i | H 
| \psi \rangle
\nonumber \\  && i \hbar {d \Phi_i \over dt} = \sum_{j (\neq i)}
{\sqrt{p_i}\langle\phi_i
\Phi_j | H | \psi \rangle - \sqrt{p_j} \langle \psi | H | \phi_j 
\Phi_i \rangle \over
p_i - p_j}
\Phi_j \nonumber
  \\
&& i \hbar {d\phi_i \over dt} = \sum_{j(\neq i)} {\sqrt{p_i}\langle 
\phi_j \Phi_i | H
| \psi \rangle - \sqrt{p_j} \langle \psi | H | \phi_i \Phi_j \rangle
\over p_i - p_j} \phi_j \nonumber \\ && \quad \quad\quad\quad\quad + 
\sqrt{p_i}\, {\rm
Re}
\langle
\phi_i
\Phi_i | H |
\psi
\rangle \phi_i
\eea
  (Note added: the asymmetry between the second and third equation can 
be avoided by
choosing a more appropriate, symmetric phase convention for the factor states
of (2)
  -- see Ph.\ts Pearle, Int.\ts J.\ts Theor.\ts Phys.\ts {\bf 18}, 489
(1979).) The ``probability resonances"
$1/(p_i - p_j)$ appearing in (3) drop out from the time derivatives 
of the reduced
density matrices,
\bea
i \hbar {d\rho_\Phi \over dt} &&:= i \hbar {d\sum p_i \Phi_i 
\Phi_i^\ast \over dt}
\nonumber \\ &&=
\sum_{i,j}\left( \sqrt{p_i}\langle \phi_i \Phi_j|H| \psi \rangle 
-\sqrt{p_j}\langle
\psi |H| \phi_j \Phi_i \rangle
\right) \Phi_i \Phi_j^\ast
\eea
and correspondingly for $\rho_\phi = \sum_i p_i \phi_i \phi_i^\ast$. Only for
vanishing interaction ($H = H_\phi + H_\Phi$) or when the
wave function happens to factorize ($p_i = \delta_{ii_0}$) does one recover the
Schr\"odinger equation, for example
\be
i \hbar {d \phi_{i_0} \over dt} = \sum_{j} \langle \phi_j |H_\phi^{(i_0)} |
\phi_{i_0} \rangle \phi_j \quad ,
\ee
where  the effective $i$-dependent Hamiltonian
$
H_\phi^{(i)} = \langle \Phi_i | H | \Phi_i \rangle
$
in our phase convention includes  the $\Phi_i$-energy (but see 
Pearle, l.c.). And
only for special initial conditions is (4) approximated by an autonomous master
equation for the density matrix.

Equations  (3) lead to a number of
phenomena that are not present in the collapse interpretation. For example,
time-reversed branchings (recombinations) may occur. If the other
branches are not known, recombinations have to be interpreted as 
irreproducible
events.

It is particularly interesting to study dynamical effects relating 
states of different
``memory". Memory states $\phi_i$ are here defined
as dynamically stable ones in the presence of a ``normal" environment 
(consider the
chiral states of a sugar molecule as a simple example). This means
\be
\langle \phi_i \Phi_j|H | \phi_{i'} \Phi_{j'} \rangle \approx 
\delta_{ii'}\langle
\Phi_j |H_\Phi^{(i)} | \Phi_{j'} \rangle
\ee
for all relevant environmental states $\Phi_j$.

Whenever an element of the Schmidt form
coincides with one of these stable states,  the time derivative of 
the corresponding
coefficient vanishes,
\be
\hbar{d\sqrt{p_i} \over dt} \approx \sqrt{p_i}\, {\rm Im} \langle 
\phi_i \Phi_i | H|
\phi_i
\Phi_i \rangle = 0 \quad ,
\ee
so this component keeps its probability. If the non-negligible other
elements are stable, too, their ``relative states" $\Phi_i$ can be 
shown by some
rearrangement of the second equation (3) to vary according to
\be
i\hbar {d\Phi_i \over dt} = \sum_{j(\neq i)} \left[\langle \Phi_j | 
H_\Phi^{(i)} |
\Phi_i
\rangle
  -  {p_j \over p_i - p_j} \langle \Phi_j |
H_\Phi^{(j)}-H_\Phi^{(i)}|\Phi_i \rangle \right] \Phi_j \quad ,
\ee
  while even the ``stable" state itself
is bound to change non-trivially in this representation:
\be
i \hbar {d\phi_i \over dt} = \sum_{j(\neq i)} {\sqrt{p_i p_j}\over 
p_i - p_j}\langle
\Phi_i |(H_\Phi^{(i)} - H_\Phi^{(j)})| \Phi_j \rangle \phi_j + p_i 
\langle \Phi_i|
H_\Phi^{(i)} | \Phi_i\rangle \phi_i \quad .
\ee
If $\Phi$ describes a macroscopic system, while $H_\Phi^{(i)}$  and 
$H_\Phi^{(j)}$
differ sufficiently, $\Phi_i$ and $\Phi_j$ must soon differ in
many degrees of freedom, and the (partial) matrix elements $\langle \Phi_i | H|
\Phi_j
\rangle$ become extremely small for $i \neq j$. The Schmidt states then tend to
stick to the dynamically stable states [3].

  If a stable state $\phi_i$ occurs as a
pure state, $p_i = 1$, its ``rate of deseparation" (or entanglement 
rate -- see Eq.\ts
(12a) of Ref.\ts 9),
\be
A = \sum_{j(\neq i)j' (\neq i)} |\langle \phi_j\Phi_{j'} |H| \phi_i 
\Phi_i \rangle
|^2 \quad ,
\ee
vanishes. This
quantity A measures the entanglement rate of initially factorizing 
systems in second
order of time. If the initial state happens to be a {\it 
superposition} of stable
states, e.g.\ts $\psi = (\phi_i + \phi_j)\Phi /\sqrt{2}$, A can be written in 
the
form $|| P_\perp (H_\Phi^{(i)}- H_\Phi^{(j)})\Phi||^2 /2$, where $P_\perp := 1-
|\Phi\rangle
\langle
\Phi |$ is the projector on to the complement space of $\Phi$.

The resonance terms of (8)
and (9) in principle distinguish the Everett version dynamically from
the collapse version of quantum theory. If two coefficients $p_i$ of 
the Schmidt
decomposition seem to intersect in time, this leads to a behavior 
similar to that
known from two crossing energy levels when an external parameter is changed. In
particular, the coefficients of two world components do {\it not} 
cross unless their
Hamiltonian matrix element vanishes exactly. Instead they repel each 
other, thereby
exchanging their states. This appears to be a very drastic effect. 
However, since the
states also exchange their memory, any observer existing in one of 
the components
will ``believe" to be the causal successor of an observer in the 
other Schmidt branch.
The event would be felt only during the (in general extremely short)
time of resonance when superpositions of memory states may form the Schmidt
representation. It is again irreproducible unless the ``other" components were
distributed in a regular way.

One should keep in mind that the dynamical equations
(3)  are based on the assumption of a time-independent cut. This may 
not be entirely
realistic for the present purpose.

The resonance terms cannot occur in the collapse
version. The latter requires nonlinear and
time direction-dependent terms in the equation of motion [10]. In Everett's
interpretation, the absence, in reality, of recombinations (or, more generally,
the approximately autonomous dynamics of different world components) 
can be derived
from the assumption of sufficiently small amplitudes for most 
``other" components,
that is, from a statistically improbable cosmological initial  condition.
\eject

\vskip .5cm \noindent
{\bf 4. INFORMATION CONCEPT IN STATISTICAL PHYSICS} \vskip.5cm

\noindent
The $\Phi$-system, say,  may be further divided
into two subsystems. Its Schmidt states may then again be expanded in 
analogy to (2):
\be
\Phi_i = \sum_\alpha 
\sqrt{q_{i\alpha}}\chi^{(1)}_{i\alpha}\chi^{(2)}_{i\alpha} \quad
.
\ee
  For {\it microscopic} sytems, no further reduction or branching 
would in general be
meaningful with respect to this decomposition, as is well known from 
many kinds of
Einstein-Podolski-Rosen phenomena [11].

Now assume that the states $\chi^{(1)}_{i\alpha}$ contain some 
dynamically stable
degrees of freedom, labeled by an index $n$,
\be
\chi^{(1)}_{i\alpha} = \sum_{nm} c^{(i\alpha)}_{nm}\chi^{(1)}_{nm} \quad ,
\ee
  where $m$ describes all other degrees of freedom. Assume further that
the memory properties $n$ do not only affect the remainder of the world,
$\chi^{(2)}$, but may also allow their ``observation" (by means of a 
von Neumann-type
interaction) by the $\phi$-system (now regarded as the observer),
\be
\phi_i \chi^{(1)}_{nm}\chi^{(2)}_{i\alpha} \to
\phi_{in}\chi^{(1)}_{nm}\chi^{(2)}_{i\alpha n} \quad ,
  \ee
with different final states of the observer, $\langle \phi_{in} 
|\phi_{in'} \rangle
\approx
\delta_{nn'}$. If, as a consequence of the ``cosmological assumption" 
mentioned at the
end of Section 3, the Everett decomposition with respect to $i$ is 
not affected by
the further decomposition with respect to $n$ (that is, $\langle \phi_{in}
|\phi_{i'n'} \rangle \approx
\delta_{nn'}\delta_{ii'}$),
the new Schmidt decomposition becomes
\be
\psi = \sum_{in} \phi_{in}\Phi^{(in)} \quad ,
\ee
with not yet normalized though orthogonal
relative states
\be
\Phi^{(in)} = \sqrt{p_i} \sum_{\alpha m}
\sqrt{q_{i\alpha}}c^{(i\alpha)}_{nm}\chi^{(1)}_{nm}\chi^{(2)}_{i\alpha 
n} \quad .
\ee
(Note added: Here one assumes that the observer states possess a sufficient
memory capacity, that is, low entropy.) The norm of these relative states is
\bea
\langle \Phi^{(in)} | \Phi^{(in)} \rangle && =  p_i \sum_{\alpha \alpha ' m}
\sqrt{q_{i\alpha}q_{i\alpha '}} c^{(i\alpha)}_{nm} c^{\ast(i\alpha 
')}_{nm} \langle
\chi_{i\alpha ' n}^{(2)} |
\chi_{i\alpha n}^{(2)}
\rangle \nonumber \\ &&\approx p_i \sum_{\alpha 
m}q_{i\alpha}|c^{(i\alpha)}_{nm}|^2
\eea
when assuming random phases. The Schmidt representation now includes the
memory states $n$,
\be
\psi \approx \sum_{in} \sqrt{  p_i \sum_{\alpha m} 
q_{i\alpha}|c_{nm}^{(i\alpha)} |^2
}
\phi_{in} \Phi_{in} \quad ,
\ee
where normalized $\Phi_{in}$ are used instead of the relative states 
$\Phi^{(in)}$.

If the components of the Everett decomposition are weighted  by their
squared norms [6], the branching ratio between $n$ and $n'$ is
\be
{\sum_{\alpha m}q_{i \alpha} |c_{nm}^{i\alpha}|^2 \over \sum_{\alpha 
m}q_{i \alpha}
|c_{n'm}^{i\alpha}|^2 } = {{\rm tr} [P_n \rho_i^{(1)}] \over {\rm tr} [P_{n'}
\rho_i^{(1)}] } \quad ,
\ee
with
\be
P_n = \sum_m |\chi_{nm}^{(1)}\rangle \langle \chi_{nm}^{(1)} |\quad 
{\rm and} \quad
\rho_i^{(1)} = \sum_\alpha q_{i \alpha} |\chi_{i\alpha}^{(1)}\rangle \langle
\chi_{i \alpha}^{(1)} | \quad .
\ee
This is equivalent to averaging over an ensemble of pure states
$\chi^{(1)}_{i\alpha}$ with  weights $q_{i\alpha}$ ($i$ fixed), 
defining in general a
large value of the entropy $S_i^{(1)} = -\sum_\alpha q_{i \alpha} {\rm ln} q_{i
\alpha}$. A quantum theory of observation along the lines suggested 
here therefore
leads to (fictitious) ensembles without introducing an {\it a priori} 
concept of
information [3,12], in contrast to the Gibbs-Jaynes concept, for 
example [13]. In the
language of d'Espagnat [14], {\it there are no proper mixtures}. The 
constraints
describing known expectation values, chosen ad hoc and taken into 
account by means of
Lagrangian multipliers by Jaynes, are here determined physically, controlled by
questions of dynamical stability.

The major problem of classical statistical mechanics -- How can ensembles be
justified? -- is thereby reversed: How can microscopic systems be 
prepared in pure
states? This is obviously achieved by correlating the states of a 
microscopic system
completely with the observer system $\phi$ and taking into account the Everett
branching (or the reduction). This process describes what in common language is
called an increase of information. In practice, it is performed by 
correlating the
micro-system with appropriate memory states of an intervening system 
(a measuring
device).  The stability of memory states leads to an ``objectivization" of the
corresponding measurement results. Microscopic systems are usually 
reduced to their
energy eigenstates by means of the interactions with the 
electromagnetic quantum field
in an expanding and absorbing universe.

Although we are led to describe physical systems by apparent ensembles of
states, the special role played by {\it canonical} ensembles is not immediately
obvious. In particular, since the $\chi^{1}$-system, say, is 
dynamically described
by an equation analogous to (4), there is in general no Hamiltonian (not even a
time-dependent one) with respect to which the canonical ensemble 
could be defined. On
the other hand, the non-existence of information beyond the mean
energy about systems containing no memory states is plausible from dynamical
considerations corresponding to Poincare's theorem concerning the 
non-analyticity of
the constants of motion. The non-separability of quantum systems may 
then be regarded
as a quantum mechanical analog of this classical theorem.
\vskip.5cm
\noindent{\bf REFERENCES}  \vskip.5cm
\noindent
1. J.\ts v.\ts Neumann, Mathematische Grundlagen der
Quantenmechanik\\ (Springer, Berlin,\ts 1932).\\
2. E.\ts P.\ts Wigner, Am.\ts J.\ts Phys. {\bf 31},\ts 6 (1963).\\
3. H.\ts D.\ts Zeh, in Proc.\ts {\bf 49}th Enrico Fermi School of 
Physics (Academic,
New York, 1972).\\
  4. H.\ts Everett, Rev.\ts Mod.\ts Phys.\ts {\bf 29}, 454 (1957); 
B.\ts C.\ts de Witt,
in Proc.\ts {\bf 49}th Enrico Fermi School of Physics (Academic, New 
York, 1972);
Physics Today {\bf 23}, 30 (1970). \\
5. H.\ts Margenau, Phil.\ts Sci.\ts {\bf 30}, 138 (1963); Ann.\ts 
Phys.\ts (N.Y.) {\bf
23}, 469 (1963); P.\ts A.\ts Moldauer, Found.\ts Phys.\ts {\bf 2}, 41 
(1972). \\
6. H.\ts D.\ts Zeh, Found.\ts Phys.\ts {\bf 1}, 69 (1970) -- reprinted in J.A.\ts
Wheeler and W.H. Zurek, Quantum Theory and Measurement (Princeton 
1983); H.\ts Primas,
Lecture Notes (Zurich, 1971, unpublished).
\\ 7. E.\ts Schmidt, Math.\ts Annalen
{\bf 63}, 433 (1907); E.\ts Schr\"odinger, Proc.\ts Cambridge 
Phil.\ts Soc.\ts {\bf
31}, 555 (1935).
\\ 8. J.\ts S.\ts
Bell, CERN report TH 1424. \\
9. 0.\ts K\"ubler and H.\ts D.\ts Zeh, Ann. Phys.\ts (N.Y.) {\bf 
76},405 (1973).\\
10. E.\ts P.\ts Wigner, in The Scientist Speculates, L.\ts I.\ts Good, ed.\ts
(Heinemann, London, 1962), p.\ts 284. \\
11. A.\ts Einstein, N.\ts Rosen, and B.\ts Podolski, Phys.\ts Rev.\ts 
{\bf 47}, 777
(1935). \\
12. M.\ts R.\ts Schafroth, Helv.\ts Phys.\ts Acta {\bf 32}, 349 (1959). \\
13. E.\ts T.\ts Jaynes,
Phys.\ts Rev.\ts {\bf 106}, 620 (1957); {\bf 108}, 171 (1957). \\
14. B.\ts d'Espagnat, Conceptual
Foundations of Quantum Theory (Benjamin, New York, 1971).

\end{document}